# Stable, highly nonlinear amorphous silicon nanowires with very low nonlinear absorption

L. Carletti*[a], C. Grillet [a,b], P. Grosse[c], B. Ben Bakir[c], S. Menezo[c], J.M. Fedeli[c], D.J. Moss [b,d], and C. Monat [a]

[a]Université de Lyon, Institut des Nanotechnologies de Lyon (INL), Ecole Centrale de Lyon, 69131 Ecully, France;
[b]Institute of Photonics and Optical Sciences (IPOS) and CUDOS, School of Physics, University of Sydney, New South Wales 2006 Australia; [d] Present Address: School of Electrical and Computer Engineering, RMIT University, Melbourne, Victoria, Australia 3001;
[c]CEA-Leti MINATEC Campus, 17 rue des Martyrs 38054 Grenoble Cedex 9, France

## ABSTRACT

The nonlinear characteristics of hydrogenated amorphous silicon nanowires are experimentally demonstrated. A high nonlinear refractive index, $n_2=1.19\times10^{-17}$ m$^2$/W, combined with a low two-photon absorption, $0.14\times10^{-11}$ m/W, resulted in a high nonlinear FOM of 5.5. Furthermore, systematic studies over hours of operational time under 2.2W of pulse peak power revealed no degradation of the optical response.

## 1. INTRODUCTION

In nowadays telecommunication networks the bandwidth demand continues to rise at an exponential rate. In order to comply with such a pressing requirement, compact on-chip architectures for all-optical signal processing that are energetically efficient and leverage the cheap CMOS manufacturing process developed for the microelectronic industry are necessary. Optical nonlinear phenomena such as four-wave mixing (FWM) [1], third harmonic generation (THG) [2,3] and self-phase modulation (SPM) [4] form the basis for such photonic nonlinear devices [5]. Many demonstrations of all-optical signal processing [6,7], wavelength conversion [8], signal monitoring [9], and signal regeneration [10] that make use of the nonlinear optical properties of semiconductor materials can be already found in the literature. Within this field, single crystal silicon-on-insulator (SOI) has been the most promising material platform. On the one hand, its mature fabrication and processing technology is fully compatible with the industrial scale production system of computer chip technology (CMOS). On the other hand, the large nonlinear refractive index coefficient, $n_2$, of crystalline silicon (c-Si) combined with nano-waveguides structures, which allow a tight confinement of light, allows to achieve a high nonlinear coefficient (Re($\gamma$) = $\omega n_2 / cA_{eff}$, where $\omega$ is the angular frequency, $n_2$ is the nonlinear refractive index and $A_{eff}$ is the waveguide effective area) exceeding 300W$^{-1}$ m$^{-1}$ around 1550nm [5]. Nevertheless, c-Si suffers from a high nonlinear absorption at telecommunication wavelengths mainly due to two-photon absorption (TPA) and the associated generation of free carriers. To quantify the impact of nonlinear losses on applications, a nonlinear figure of merit is introduced. This is defined as FOM = $n_2 /(\lambda\, \beta_{TPA})$ where $\lambda$ is the wavelength and $\beta_{TPA}$ is the TPA coefficient. For c-Si, the achievable FOMs in the telecom band are in the range of 0.3-0.5 [5]. Such a low FOM has impeded the development of efficient nonlinear devices based on c-Si and operating in the telecom band interval around the 1550nm wavelength, e.g. for positive net gain amplification [11].

In order to overcome the TPA impairment of c-Si in the C-band without losing the technological advantages of a CMOS compatible platform, amorphous hydrogenated silicon (a-Si:H) has recently been suggested as a possible alternative material [12]. Its bandgap energy (about 1.7eV) larger than the one of c-Si (1.12ev) reduces the strength of the TPA process around 1550nm and thus potentially increases the achievable FOM. Experimental demonstrations showed the possibility to increase the FOM as high as 1 [13,14] or 2 [15,16], allowing the achievement of very high parametric gains of over +26dB over the C-band and, more recently, the realization of ultralow-power all-optical processing of high-speed data signals [17]. However, the material fabricated and used in the latest demonstrations, [16], has proved to have an unstable nonlinear optical response even over short operation time (around tens on minutes). This stability issue has to be solved in order to seriously consider a-Si:H as a material of choice replacing c-Si for nonlinear chip-based photonic devices.. In this work a-Si:H nanowires exhibiting simultaneously a high nonlinear coefficient, Re($\gamma$), and a high nonlinear FOM are reported. Furthermore, the fabricated samples show stable nonlinear optical response over several hours of operation time. The nonlinear coefficient and the TPA coefficient, $\beta_{TPA}$, are determined by experimental measurements of the nonlinear transmission and self-phase modulation induced spectral broadening fit to numerical simulations and standard theory results [18].

## 2. FABRICATION AND CHARACTERIZATION SETUP

The a:Si-H waveguides were fabricated onto a SiO$_2$/Si wafer using a 200mm CMOS pilot line at CEA-LETI, France. The a:Si-H film was deposited by plasma enhanced chemical vapor deposition (PECVD) at 350°C on 1.7µm oxide deposited on a bulk wafer. After deposition of a silica hard mask, two steps of 193nm DUV lithography and HBr silicon etching were used to define grating couplers that were well aligned with serpentine waveguides with varying lengths (2.28cm to 11.04cm). The fabricated nanowires have a thickness of 220nm and a width of about 500nm. Finally, a 500nm high density plasma oxide deposition completed the fabrication in order to perfectly fill the gaps between waveguides. The cross-section of the fabricated waveguide is shown in Fig. 1(a). The group velocity dispersion for the TE mode confined within a 500nm×220nm nanowire was calculated with FEMSIM, providing an anomalous second-order dispersion parameter $\beta_2$=-4.2×10$^{-25}$ s$^2$/m at $\lambda$=1550nm.

In order to experimentally measure the linear and nonlinear transmission of the fabricated a-Si:H nanowires, the setup depicted in Fig. 1(b) is used. The light source is a passively mode-locked fiber laser emitting near transform limited pulses of about 1.8ps duration at a repetition rate of 20MHz and at a wavelength tunable around 1550nm. Using a polarization controller, the TE polarization of the input signal was selected and coupled into the nanowires via in-plane gratings. The estimated single mode fiber to waveguide coupling loss of the couplers was about 12dB and 10.5dB per entry and exit, respectively, which is higher than designed due to the grating couplers not being optimized.

To determine the nonlinear parameters of the waveguides, two series of self-phase modulation (SPM) measurements were performed on a 2.28cm and a 4.72cm long nanowire, respectively, with a coupled peak power up to 2.5W. The output spectrum was recorded as a function of input power.

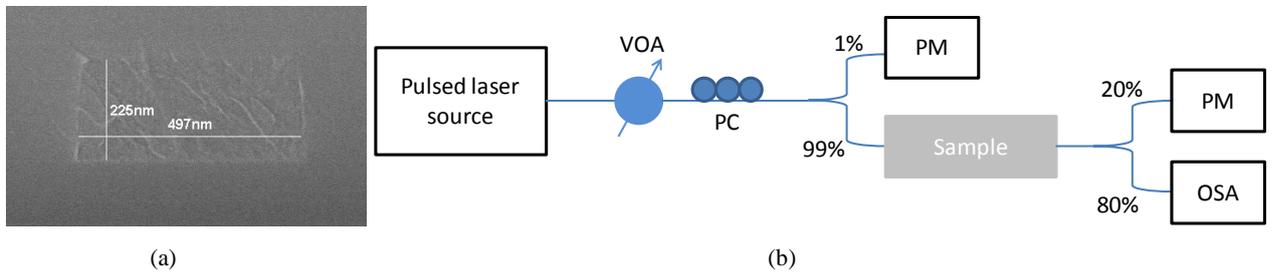

(a) (b)

Figure 1. (a) Cross-section image obtained by scanning electron microscopy of the fabricated a-Si:H nanowires embedded into silica. (b) Schematic of the experimental setup used for the characterization of the a-Si:H nanowires. The pulsed laser source output passes through a variable optical attenuator (VOA) and a polarization controller (PC). A 1:99 splitter allows coupling light simultaneously to the sample and to a power meter (PM). At the sample output another splitter (20:80) is used to direct light to a second PM and to an optical spectrum analyzer (OSA).

## 3. RESULTS AND DISCUSSION

### 3.1 Nonlinear characterization

Using the cut-back method on serpentine nanowire waveguides with lengths ranging from 2.28cm to 11.04cm, the linear propagation loss coefficient, $\alpha$, of the TE mode was measured to be 4.5 dB/cm or equivalently 0.52 cm$^{-1}$ around 1550nm. In order to experimentally determine the nonlinear characteristic of the fabricated a-Si:H nanowires, a study on the nonlinear transmission and SPM phenomenon were performed versus input power of the pulsed laser signal on two waveguide lengths (2.28cm and 4.72cm). The transmitted spectra on Fig. 2 are recorded when varying the input pulse peak power from 0.3W up to 2.5W. The observed spectral broadening as the peak power is increased is a clear signature of the SPM phenomenon experienced by the optical pulses in the nonlinear waveguides. Due to the limited bandwidth of the in-plane grating couplers used to couple light in and out of the waveguides, the observed spectral broadening is influenced by the grating transfer function for peak powers bigger than about 1.8W. The 3dB bandwidth of these grating couplers was measured to be about 35nm and centered at a wavelength of 1550nm. The filtering effect of the grating coupler at the output can be better seen in the measured transmitted power shown in Fig. 3.

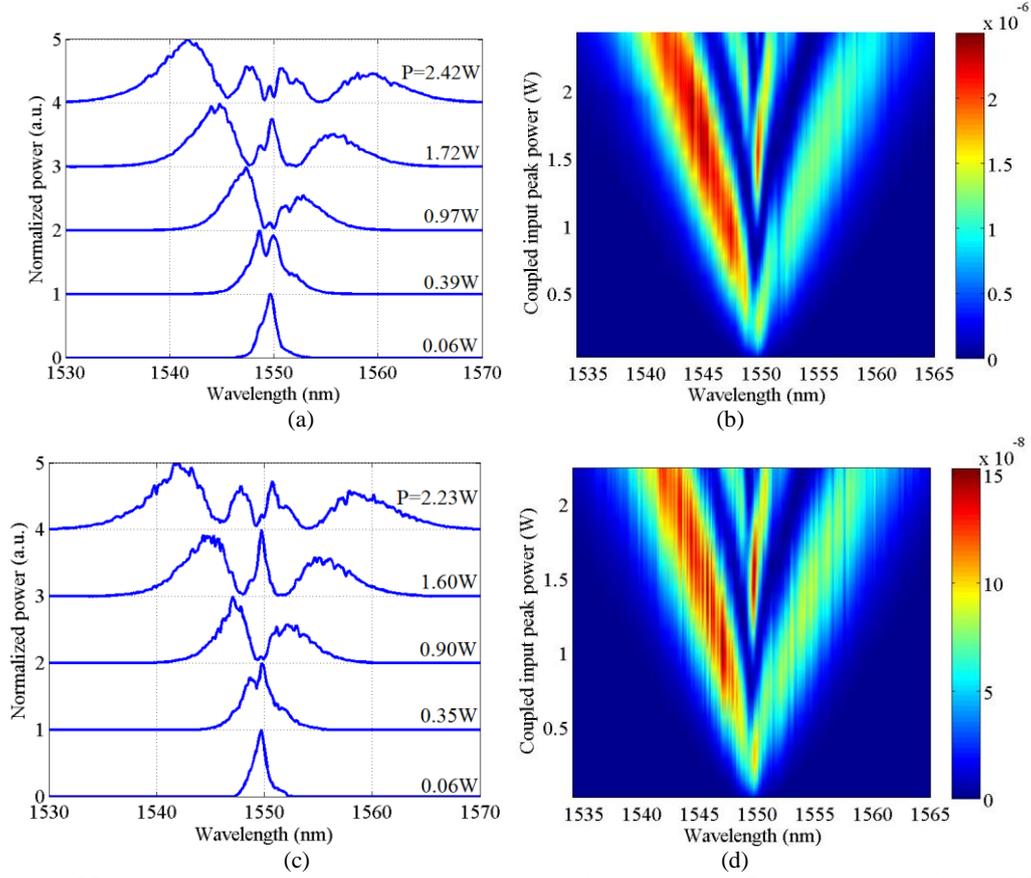

Figure 2. (a)-(d) Measured output spectrum as a function of the coupled peak power, P, for nanowires of length 2.78cm, (a) and (b), and 4.72cm, (c) and (d). The color-scale depicts the output power measured by the optical spectrum analyzer in units of W.

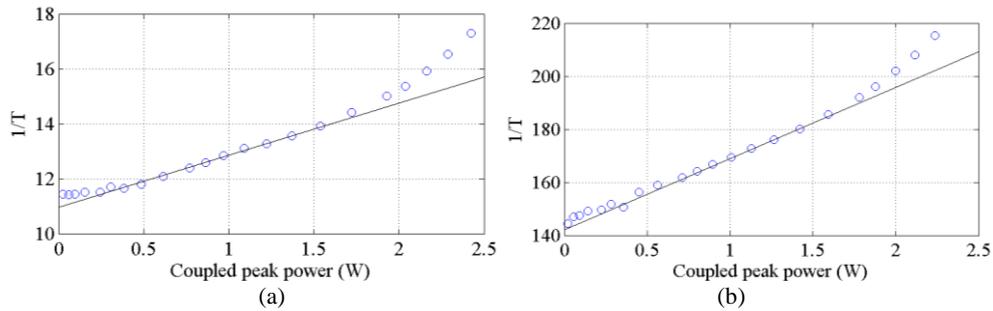

Figure 3. (a)-(b) Inverse of the measured transmission as a function of the coupled peak power for nanowires of length 2.78cm and 4.72cm respectively. The experimental data (blue circles) are shown along with the linear fit (black line).

In Fig. 3 the inverse of the measured transmitted power, 1/T, as a function of the coupled peak power is reported. Under the assumption of the presence of TPA, the inverse of the transmission, 1/T, should linearly increase as the input peak power increases [19]. However, at high peak power, the broadened output spectrum is filtered by the pass-bandwidth of the grating coupler and thus a drop of T, or a strong increase of 1/T, is observed for peak powers greater than about 1.8W. Thus, the linear behavior of 1/T is assumed valid only at low input powers. The nonlinear losses, associated with the imaginary part of the nonlinear coefficient, Im($\gamma$), can be estimated by a linear fit of 1/T as a function of the input peak power using equation (1):

$$\frac{1}{T} = \frac{P(0)}{P(L)} = 2\,\text{Im}(\gamma) L_{eff} e^{\alpha L} P(0) + e^{\alpha L} \qquad (1)$$

where $L$ is the physical nanowaveguide length, P(0) is the coupled peak power, P($L$) is the peak power of the pulse at the end of the waveguide, $\alpha$ is the linear loss coefficient previously measured by cut-back method and $L_{eff} = (1-e^{\alpha L})/\alpha$ is the effective length of the waveguide. The imaginary part of the nonlinear coefficient, Im($\gamma$), includes all the nonlinear loss phenomena. In our experiment, the pulse duration, i.e. 1.8ps, and the repetition rate of the pulsed laser, i.e. 20MHz, are small enough to, a priori, neglect the generation of free carriers and the associated free carrier absorption. Thus, the nonlinear losses can be interpreted as only generated by the TPA mechanism. In this hypothesis, we can directly relate the TPA coefficient to Im($\gamma$) through the formula Im($\gamma$) = $\beta_{TPA}$ / 2$A_{eff}$ where $A_{eff}$ is the effective area of the fundamental TE mode in the nanowire assumed to be equal to 0.07 µm$^2$. By using Eq. (1) to fit the experimental data, we extract Im($\gamma$) = (10±5%) W$^{-1}$m$^{-1}$ and, by making use of the above relation, the TPA coefficient, $\beta_{TPA}$, is estimated to be equal to (0.14×10$^{-11}$±5%) m/W. The uncertainty on the extracted parameters is mainly due to the uncertainty on the coupling loss of both the input and output grating couplers that could not be exactly measured [20]. However, this error could be somewhat reduced through comparing the nonlinear data obtained from launching the signal from either side [18]. Furthermore, in Fig. 2 it is noted that there is no blue shift of the output spectra as the coupled peak power is increased. This blue shift is generally a signature of free carriers generated by TPA over the pulse duration and has been commonly observed in c-Si waveguides for near-infrared picosecond pulses exhibiting similarly high nonlinear phase shifts [21].

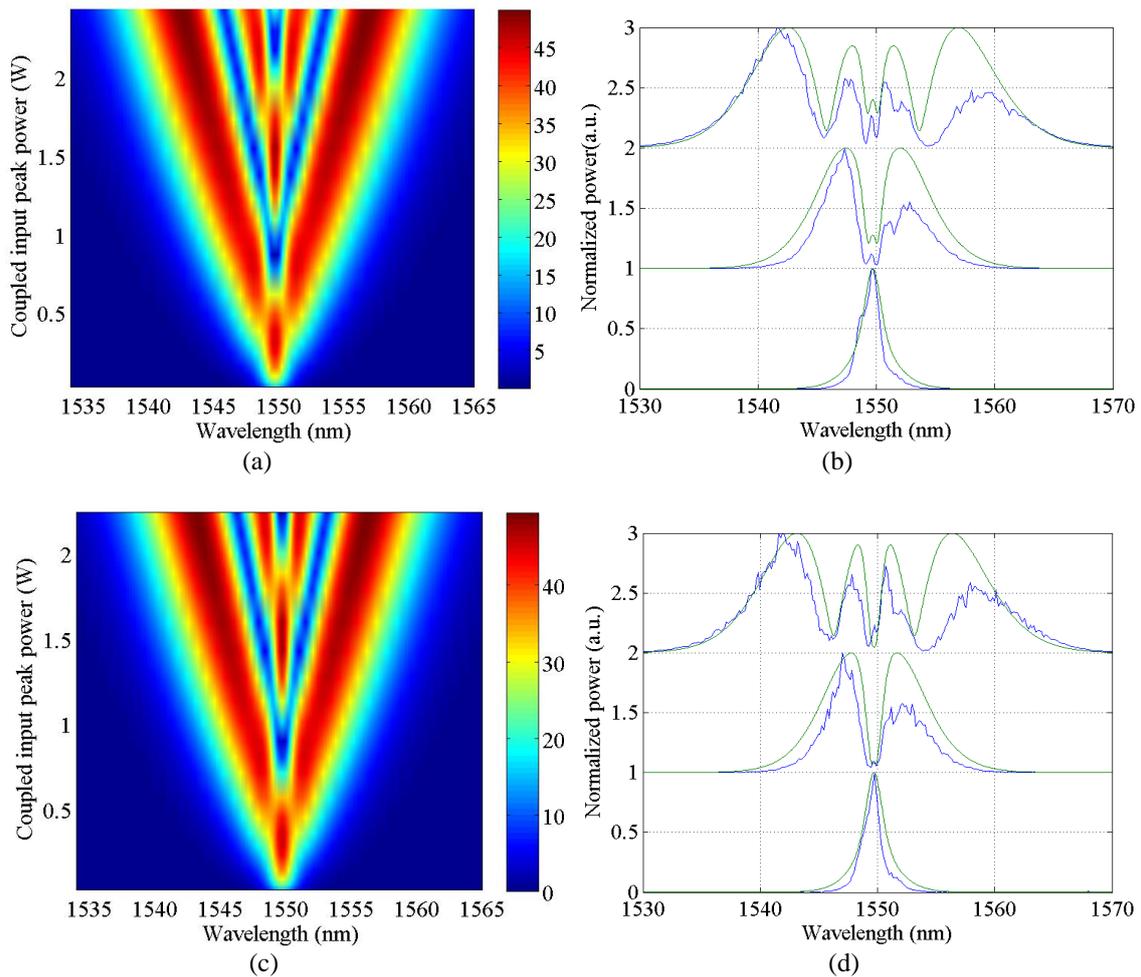

Figure 4. Output spectrum as a function of the input peak power obtained from SSFM simulations. (a) and (b) are obtained for a simulated nanowire length of 2.78cm, while (c) and (d) for a nanowire length of 4.72cm. The color-scale represents normalized power in linear scale. In (b) and (d) the experimental measured spectra (blue lines) are compared to simulation results (greed lines) for coupled peak powers of 0.14W, 0.9W and 2.24W from the bottom to the top of the plots respectively.

The nonlinear Schrödinger equation that describes the propagation of picosecond optical pulses through waveguides taking into account both linear and nonlinear phenomena is solved by using the split-step Fourier method (SSFM). This equation, by neglecting high-order dispersion effects other than GVD, has the form [22]

$$\frac{\partial A}{\partial z} + \frac{i\beta_2}{2}\frac{\partial^2 A}{\partial t^2} = -\frac{\alpha}{2}A + i\gamma|A|^2 A \qquad (2)$$

where $A$ is the slowly varying amplitude of the pulse envelope, $z$ is the propagation axis, and $\gamma$ is the complex nonlinear coefficient, i.e. $\gamma = \mathrm{Real}(\gamma) + i\cdot\mathrm{Im}(\gamma)$. The physical parameters ($\beta_2$, $\alpha$, and $\gamma$) adopted to solve Eq. (2) are the same of the fabricated nanowires. The results for the two waveguide lengths are shown in Fig. 4.

By comparing Figs. 4 and 2 a good agreement between the experimental data and the simulation results is observed when using $\mathrm{Real}(\gamma)= 690\,\mathrm{W}^{-1}\mathrm{m}^{-1}$ and $\mathrm{Im}(\gamma)= 10\,\mathrm{W}^{-1}\mathrm{m}^{-1}$. In Fig. 5, the nonlinear phase-shift as a function of the coupled power is shown. From the experimental results, this phase shift is deduced from the measurement of the spectral broadening obtained by calculating the root mean square spectral width of the recorded spectra [22]. The experimental results are fit to the results obtained by SSFM and a good agreement between the different data is observed. By comparing experiment and simulation results for the two waveguides, the deduced real part of the nonlinear coefficient is $\mathrm{Real}(\gamma)=690\,\mathrm{W}^{-1}\mathrm{m}^{-1}$ that corresponds to a nonlinear refractive index $n_2 = (1.19 \times 10^{-17} \pm 5\%)\,\mathrm{m}^2/\mathrm{W}$.

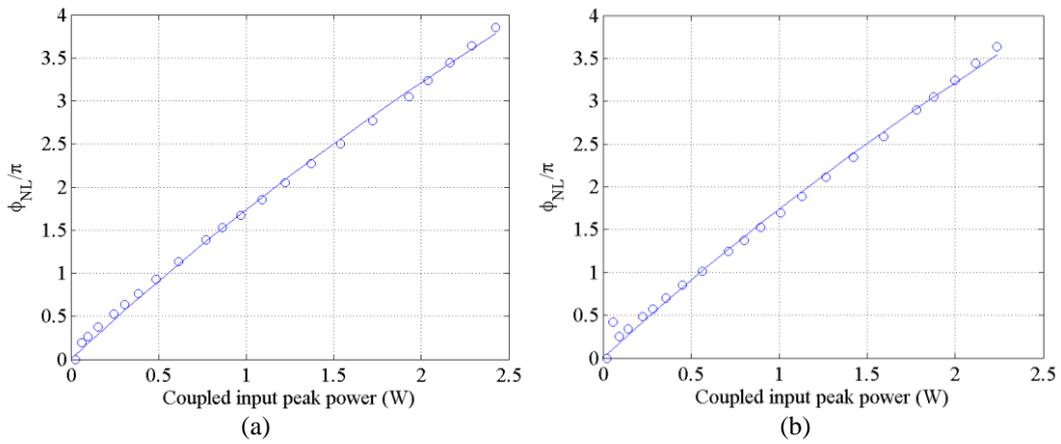

Figure 5. (a)-(b) Nonlinear phase shift, $\phi_{NL}$, as a function of the input peak power for a nanowire length of 2.28cm and 4.72cm respectively. Values deduced from experiments (blue circles) are compared to simulation results (blue line).

Finally the nonlinear FOM is estimated to be $5.5\pm0.3$, which is an order of magnitude higher than c-Si and more than double of the FOM obtained by other groups [13,15,16]. Table 1 reports the main results found in the literature. As it can be seen, high nonlinear coefficients or low TPA were already demonstrated but the overall FOM in the C-band has always been poor due to either a high TPA [13,15,16,20] or a low nonlinear refractive index coefficient [14,23]. Thus, our reported results on a-Si:H nanowires represent an important advance in the development of this material platform thanks to the achievement of both a high nonlinear parameter and a high FOM.

Table 2. Summary of the state-of-the-art achievable nonlinear characteristics that have been experimentally determined for a-Si:H waveguides.

|  | This work | [15,16] | [20] | [23] | [14] | [13] |
|---|---|---|---|---|---|---|
| $n_2$ ($10^{-17}$ m$^2$/W) | 1.19 | 1.3 | 4.2 | 0.05 | 0.3 | 7.43 |
| Re($\gamma$) (W$^{-1}$m$^{-1}$) | 690 | 770 | 2000 | 35 | N/A | N/A |
| $\beta_{TPA}$ (cm/GW) | 0.14 | 0.392 | 4.1 | 0.08 | 0.2 | 4 |
| FOM | 5.5±0.3 | 2.2±0.4 | 0.66±0.3 | 0.4 | 0.97 | 1.1 |

### 3.2 Stability analysis

Another open problem reported for the a-Si:H material platform was the degradation of the nonlinear optical response over short (few tens of minutes) operational time [16]. Thus the stability of our material was also tested by recording the output spectrum with a fixed coupled peak pulse power, 2.25W, which corresponds to about $3GW/cm^2$, every two minutes over hours of operational time. The results are reported in Fig 6.

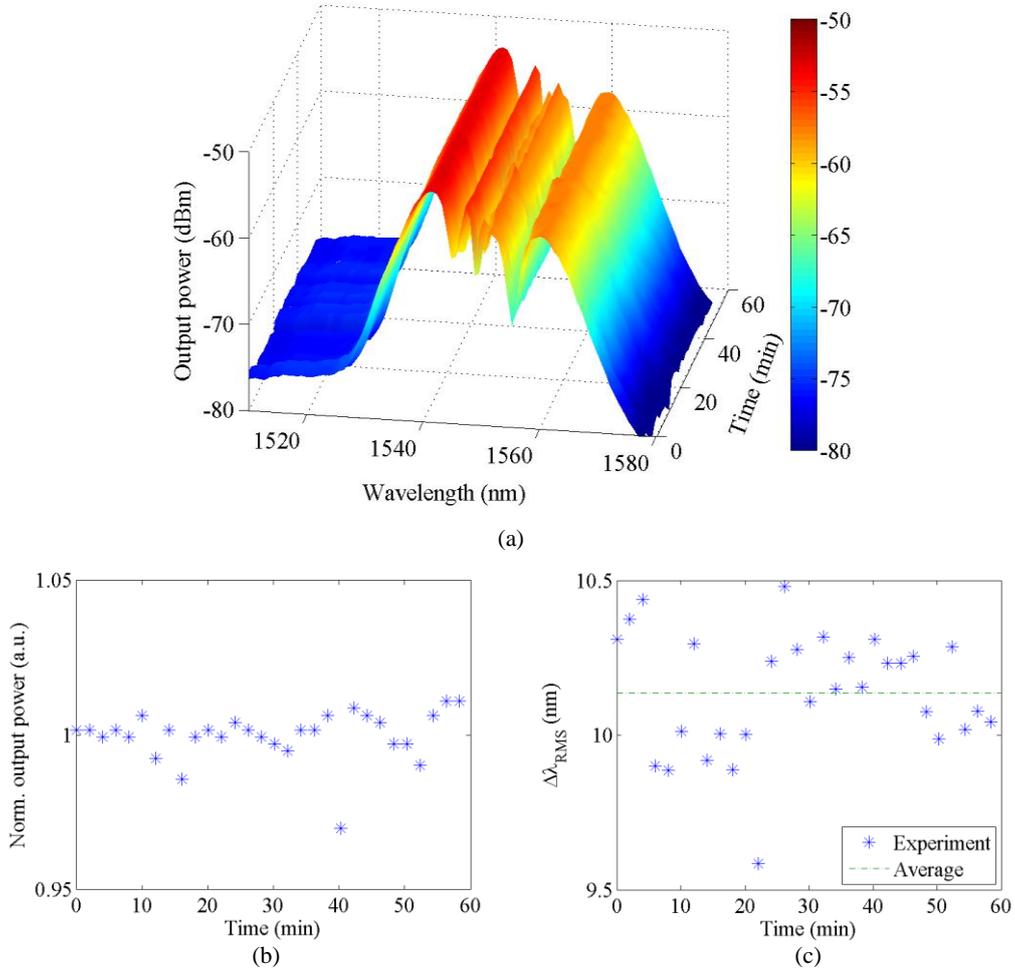

Figure 6. (a) Evolution of the output spectrum over time at a coupled pulse peak power of 2.25W for the 2.28cm long nanowire. (b) Normalized output power as a function of time. The output power is normalized to the mean value of output power recorded during the experiment (94nW). (c) RMS of the spectral broadening as a function of time. The average spectral broadening during the experiment, ~10nm, is represented by the green dashed line.

From the measured power spectra it is observed that the SPM broadening effect remains practically unchanged. As shown in Figs. 6(b)-6(c), random fluctuations of the measured RMS broadening can be attributed to small variations in the input-output coupling. Furthermore, no degradation at all in the parameters was observed over the entire course of our experiments, which at times were conducted at peak powers of up to 3W (about $4GW/cm^2$). These results are obtained with a 20 MHz repetition rate laser and the material may behave differently at higher repetition rates or under CW excitation. It is noted, though, that this material has also exhibited stable operation in the context of a hybrid integrated CW laser [24].

## 4. CONCLUSION

The nonlinear properties of fabricated a-Si:H nanowires have been experimentally measured. The high Kerr nonlinear coefficient measured, $n_2 = (1.19 \times 10^{-17} \pm 5\%)$ m$^2$/W, combined with a low TPA, $(0.14 \pm 5\%)$ cm/GW, yielded a nonlinear FOM of about 5.5 in the C-band, which is one order of magnitude bigger than the one obtained for c-Si. Furthermore, no degradation of the material optical response was found over hours of operational time at a coupled peak power of about 2.2W and a repetition rate of 20 MHz. These findings, in addition to the compatibility with the CMOS industry technology, show the great potential of the a-Si:H material platform for the conception of nonlinear photonic devices. Such components will be essential for developing efficient all-optical on-chip architecture able to efficiently satisfy the increasing demand of bandwidth for future telecommunication networks.

## REFERENCES

[1] C. Monat, M. Ebnali-Heidari, C. Grillet, B. Corcoran, B. J. Eggleton, T. P. White, L. O'Faolain, J. Li, and T. F. Krauss, "Four-wave mixing in slow light engineered silicon photonic crystal waveguides.," *Optics express* **18**(22), 22915–22927 (2010).

[2] C. Monat, C. Grillet, B. Corcoran, D. J. Moss, B. J. Eggleton, T. P. White, and T. F. Krauss, "Investigation of phase matching for third- harmonic generation in silicon slow light photonic crystal waveguides using Fourier optics," 6831–6840 (2010).

[3] B. Corcoran, C. Monat, C. Grillet, D. J. Moss, B. J. Eggleton, T. P. White, L. O. Faolain, and T. F. Krauss, "Green light emission in silicon through slow-light enhanced third-harmonic generation in photonic- crystal waveguides" (2009).

[4] C. Monat, B.Corcoran, C. Grillet, D. J. Moss, B.J.Eggleton, T.White, and T.Krauss, "Slow light enhanced nonlinear optics in dispersion engineered slow-light silicon photonic crystal waveguides", IEEE Journal of Selected Topics in Quantum Electronics (JSTQE) special issue on Silicon Photonics, 16 (1) 344-356 (2010).

[5] J. Leuthold, C. Koos, and W. Freude, "Nonlinear silicon photonics," *Nature Photonics* **4**(July), 535–544 (2010).

[6] H. Ji, M. Galili, H. Hu, M. Pu, L. K. Oxenløwe, K. Yvind, J. M. Hvam, and P. Jeppesen, "1 . 28-Tb / s Demultiplexing of an OTDM DPSK Data Signal Using a Silicon Waveguide," *IEEE Photonics Technology Letters* **22**(23), 1762–1764 (2010).

[7] F. Li, M. Pelusi, D.-X. Xu, a. Densmore, R. Ma, S. Janz, and D. J. Moss, "Error-free all-optical demultiplexing at 160Gb/s via FWM in a silicon nanowire," *Optics Express* **18**(4), 3905 (2010).

[8] B. G. Lee, S. Member, A. Biberman, A. C. Turner-foster, M. A. Foster, M. Lipson, S. Member, A. L. Gaeta, and K. Bergman, "Demonstration of Broadband Wavelength Conversion at 40 Gb / s in Silicon Waveguides," *IEEE Photonics Technology Letters* **21**(3), 182–184 (2009).

[9] B. Corcoran, C. Monat, M. Pelusi, C. Grillet, T. P. White, L. O'Faolain, T. F. Krauss, B. J. Eggleton, and D. J. Moss, "Optical signal processing on a silicon chip at 640Gb/s using slow-light.," *Optics Express* **18**(8), 7770–7781 (2010).

[10] Reza Salem, et al.., "Signal regeneration using low-power four-wave mixing on silicon chip," *Nature Photonics* **2**, 35–38 (2008).

[11] M. a Foster, A. C. Turner, J. E. Sharping, B. S. Schmidt, M. Lipson, and A. L. Gaeta, "Broad-band optical parametric gain on a silicon photonic chip," *Nature* **441**, 960–963 (2006).

[12] K. Ikeda, Y. Shen, and Y. Fainman, "Enhanced optical nonlinearity in amorphous silicon and its application to waveguide devices.," *Optics Express* **15**(26), 17761–17771 (2007).

[13] K.-Y. Wang and A. C. Foster, "Ultralow power continuous-wave frequency conversion in hydrogenated amorphous silicon waveguides.," *Optics Letters* **37**(8), 1331–1333 (2012).

[14] S. Suda, K. Tanizawa, Y. Sakakibara, T. Kamei, K. Nakanishi, E. Itoga, T. Ogasawara, R. Takei, H. Kawashima, et al., "Pattern-effect-free all-optical wavelength conversion using a hydrogenated amorphous silicon waveguide with ultra-fast carrier decay.," *Optics letters* **37**(8), 1382–1384 (2012).

[15] B. Kuyken, S. Clemmen, S. K. Selvaraja, W. Bogaerts, D. Van Thourhout, P. Emplit, S. Massar, G. Roelkens, and R. Baets, "On-chip parametric amplification with 26.5 dB gain at telecommunication wavelengths using CMOS-compatible hydrogenated amorphous silicon waveguides.," *Optics letters* **36**(4), 552–554 (2011).

[16] B. Kuyken, H. Ji, S. Clemmen, S. K. Selvaraja, H. Hu, M. Pu, M. Galili, P. Jeppesen, G. Morthier, et al., "Nonlinear properties of and nonlinear processing in hydrogenated amorphous silicon waveguides.," *Optics express* **19**(26), B146–53 (2011).

[17] K.-Y. Wang, K. G. Petrillo, M. a Foster, and A. C. Foster, "Ultralow-power all-optical processing of high-speed data signals in deposited silicon waveguides.," *Optics express* **20**(22), 24600–24606 (2012).

[18] C. Grillet, L. Carletti, C. Monat, P. Grosse, B. Ben Bakir, S. Menezo, J. M. Fedeli, and D. J. Moss, "Amorphous silicon nanowires combining high nonlinearity, FOM and optical stability.," *Optics Express* **20**(20), 22609–22615 (2012).

[19] H. K. Tsang, R. V. Penty, I. H. White, R. S. Grant, W. Sibbett, J. B. D. Soole, H. P. LeBlanc, N. C. Andreadakis, R. Bhat, et al., "Two-photon absorption and self-phase modulation in InGaAsP/InP multi-quantum-well waveguides," *Journal of Applied Physics* **70**(7), 3992 (1991).

[20] K. Narayanan and S. F. Preble, "Optical nonlinearities in hydrogenated-amorphous silicon waveguides.," *Optics Express* **18**(9), 8998–9005 (2010).

[21] E. Dulkeith, Y. a Vlasov, X. Chen, N. C. Panoiu, and R. M. Osgood, "Self-phase-modulation in submicron silicon-on-insulator photonic wires.," *Optics express* **14**(12), 5524–5534 (2006).

[22] G. P. Agrawal, *Nonlinear Fiber Optics* (2001).

[23] Y. Shoji, T. Ogasawara, T. Kamei, Y. Sakakibara, S. Suda, K. Kintaka, H. Kawashima, M. Okano, T. Hasama, et al., "Ultrafast nonlinear effects in hydrogenated amorphous silicon wire waveguide.," *Optics Express* **18**(6), 5668–5673 (2010).

[24] K. Narayanan, A. W. Elshaari, and S. F. Preble, "Broadband all-optical modulation in hydrogenated-amorphous silicon waveguides.," *Optics Express* **18**(10), 9809–9814 (2010).